\documentclass[10pt,conference,twocolumn,]{IEEEtran}
\IEEEoverridecommandlockouts

\usepackage{amsmath,amssymb,amsfonts}
\usepackage{cite}
\usepackage{graphicx}
\usepackage{epstopdf}

\usepackage{booktabs}
\usepackage{graphicx}
\usepackage{textcomp}
\usepackage{xcolor}
\usepackage{times}
\usepackage{subfigure}         
\usepackage{amssymb,amsmath}
\usepackage{acronym}  
\usepackage{balance}

\usepackage{bm}         
\usepackage{algorithm} 
\usepackage{algorithmic}

\usepackage{lipsum}
\usepackage{stfloats}

\usepackage{url}

\usepackage{mathrsfs}

\usepackage{amsmath}
\allowdisplaybreaks[4]

\newtheorem{proposition}{\bf Proposition}

\newtheorem{lemma}{\bf Lemma}
\newtheorem{corollary}{\bf Corollary}

\acrodef{lis}[LIS]{large intelligent surface}

\acrodef{ofdm}[OFDM]{orthogonal frequency division multiplexing}%
\acrodef{miso-ofdm}[MISO-OFDM]{multi-input single-output orthogonal frequency division multiplexing}%

\acrodef{ris}[RIS]{reconfigurable intelligent surface}%

\acrodef{irs}[IRS]{intelligent reflecting surface}%

\acrodef{qos}[QoS]{quality of service}%
\acrodef{idft}[IDFT]{inverse discrete Fourier transform}%
\acrodef{dft}[DFT]{discrete Fourier transform}%
\acrodef{cp}[CP]{cyclic prefix}%
\acrodef{csi}[CSI]{channel state information}%
\acrodef{awgn}[AWGN]{additive white Gaussian noise}%

\acrodef{qcqp}[QCQP]{quadratically constrained quadratic program}%
\acrodef{qp}[QP]{quadratic program}%
\acrodef{bs}[BS]{base station}%
\acrodef{qos}[QoS]{quality of service}%
\acrodef{ue}[UE]{user equipment}%
\acrodef{snr}[SNR]{signal-to-noise ratio}%
\acrodef{mmwave}[mmWave]{millimeter-wave}%

\acrodef{rf}[RF]{radio frequency}%

\acrodef{cap-mimo}[CAP-MIMO]{continuous-aperture MIMO}%

\acrodef{sinr}[SINR]{signal-to-interference-plus-noise ratio}%

\acrodef{ser}[SER]{symbol error rate}%
\acrodef{mimo}[MIMO]{multiple-input multiple-output}%

\acrodef{ace}[ACE]{adaptive cross-entropy}%
\acrodef{wsr}[WSR]{weighted sum-rate}%
\acrodef{udn}[UDN]{ultra-dense network}%

\def\BibTeX{{\rm B\kern-.05em{\sc i\kern-.025em b}\kern-.08em
		T\kern-.1667em\lower.7ex\hbox{E}\kern-.125emX}}

\begin{document}
	\title{Pattern-Division Multiplexing for Continuous-Aperture MIMO }
	\author{
		\vspace{0.2cm}
		\IEEEauthorblockN{
			Zijian~Zhang
			and
			Linglong~Dai}
		
		\IEEEauthorblockA{\IEEEauthorrefmark{0}
			Beijing National Research Center for Information Science and Technology (BNRist)\\
			Department of Electronic Engineering,
			Tsinghua University,
			Beijing 100084, China\\
			Email: zhangzj20@mails.tsinghua.edu.cn, daill@tsinghua.edu.cn
			\vspace*{-1.5em}
		}
	}
	\maketitle

\begin{abstract}
	In recent years, continuous-aperture multiple-input multiple-output (CAP-MIMO) is reinvestigated to achieve improved communication performance with limited antenna apertures. Unlike the classical MIMO composed of discrete antennas, CAP-MIMO has a continuous antenna surface, which is expected to generate any current distribution (i.e., pattern) and induce controllable spatial electromagnetic waves. In this way, the information can be modulated on the electromagnetic waves, which makes it promising to approach the ultimate capacity of finite apertures. The pattern design for CAP-MIMO is the key factor to determine the communication performance, but it has not been well studied in the literature. In this paper, we propose the pattern-division multiplexing to design the patterns for CAP-MIMO. Specifically, we first derive the system model of a typical multi-user CAP-MIMO system, which allows us to formulate the sum-rate maximization problem. Then, we propose a general pattern-division multiplexing technique to transform the design of continuous pattern functions to the design of their projection lengths on finite orthogonal bases. Based on this technique, we further propose a pattern design scheme to solve the formulated sum-rate maximization problem. Simulation results show that, the sum-rate achieved by the proposed scheme is about 260\% higher than that achieved by the benchmark scheme. 
\end{abstract}

\begin{IEEEkeywords}
	Continuous-aperture MIMO, electromagnetic information theory, pattern-division multiplexing.
\end{IEEEkeywords}
\section{Introduction}\label{sec:intro}
From 3G to 5G, the system performance of wireless communications has been greatly improved by the use of \ac{mimo} \cite{Andrews'16}. In recent years, in order to break the performance limit of conventional \ac{mimo} with limited antenna apertures \cite{HuangHu'20}, as an ultimate \ac{mimo} structure with infinitely dense antennas, \ac{cap-mimo}, which is also called as holographic \ac{mimo} \cite{Sanguinetti'21} or \ac{lis} \cite{Decarli'21}, is reinvestigated for wireless communications. Unlike the classical \ac{mimo} composed of multiple discrete antennas, \ac{cap-mimo} takes the form of a spatially-continuous electromagnetic surface \cite{HuangHu'20}, which can generate any current distribution and induce controllable spatial electromagnetic waves \cite{Yurduseven'18}. In this way, the information for receivers can be directly modulated on the spatial electromagnetic waves, thus it is promising to achieve the ultimate capacity of limited apertures.
	
The patterns, i.e., the current distributions on the continuous \ac{cap-mimo} aperture, are the key factors determining the performance of \ac{cap-mimo} \cite{Marzetta'20}. To support multi-stream transmissions, it is necessary for \ac{cap-mimo} to adopt a series of distinguishable patterns to carry different symbols. To achieve this, most existing works have directly adopted the patterns generated by the given special functions \cite{Sanguinetti'21,Decarli'21,HuangHu'20}. For example, the authors in \cite{Sanguinetti'21} considered a near-field line-of-sight scenario with a couple of linear-aperture \ac{cap-mimo} transceivers. Then, a wavenumber-division multiplexing scheme was proposed to directly generate orthogonal patterns by Fourier basis functions. In this way, the transmitted symbols are modulated on different wavenumbers of radiated electromagnetic waves respectively, thus these symbols become distinguishable, which is similar to the conventional frequency-division multiplexing.
	
Despite the existing schemes can improve the communication performance of \ac{cap-mimo}, most of these schemes are heuristic and can only be applied to some special scenarios, such as single receiver, near field, and line-of-sight transmissions \cite{Sanguinetti'21,Decarli'21,HuangHu'20}. To support \ac{cap-mimo} in general communication scenarios, it is essential to design the patterns flexibly according to the real-time channel state information. Unfortunately, according to the best of our knowledge, the research on such a general pattern design scheme for \ac{cap-mimo} has not been well studied in the literature.
	
To fill in this gap, in this paper, we propose the pattern-division multiplexing technique. Specifically, we first derive the system model including the capacity and power constraint of \ac{cap-mimo}, which allows us to formulate the sum-rate maximization problem. Then, we propose the general pattern-division multiplexing to flexibly design the \ac{cap-mimo} patterns. The key idea is to use series expansion to project the continuous pattern functions onto an orthogonal basis space, thus the design of continuous pattern functions is transformed to the design of their projection lengths on finite orthogonal bases. Finally, based on the proposed pattern-division multiplexing, a pattern design scheme is proposed to solve the formulated sum-rate maximization problem. Simulation results show that, the patterns designed by the proposed scheme are almost mutually orthogonal, the sum-rate achieved by the proposed scheme is about 260\% higher than that achieved by the existing wavenumber-division multiplexing scheme \cite{Sanguinetti'21}. 
	
The rest of this paper is organized as follows. Section \ref{sec:sys} introduces the system model of \ac{cap-mimo} and formulates the sum-rate maximization problem. The pattern-division multiplexing technique, as well as the pattern design scheme, are proposed in Section \ref{sec:Alg}. Simulation results are provided in Section \ref{sec:NSR} to validate the effectiveness of the proposed scheme and evaluate the sum-rate performance of \ac{cap-mimo}. Finally, conclusions are drawn in Section \ref{sec:con}.

	
\section{System Model and Problem Formulation}\label{sec:sys}

\subsection{System model}
	\begin{figure*}[!t]
		\centering
		\includegraphics[width=7.1in]{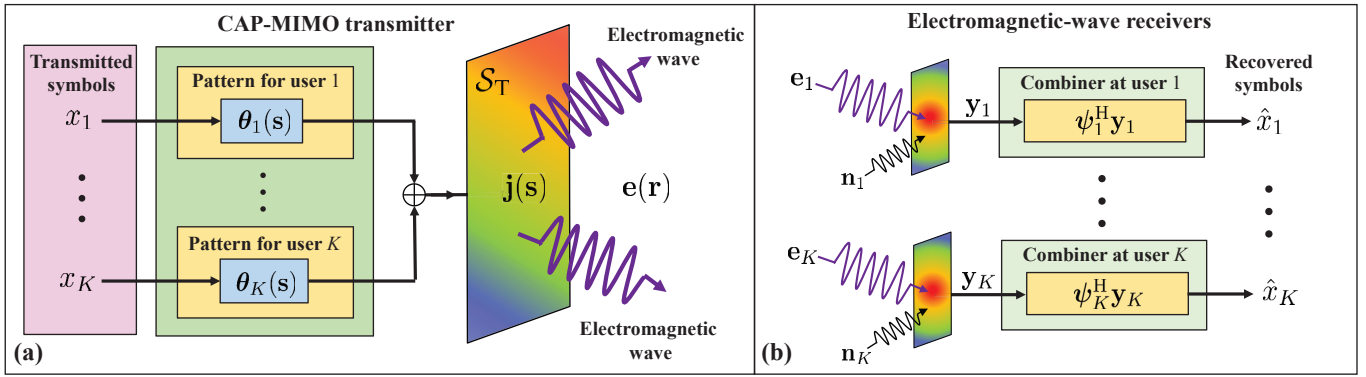}
		\caption{(a) illustrates a \ac{cap-mimo} transmitter, which has a continuous antenna aperture for radiating information-carrying electromagnetic waves \cite{HuangHu'20}. (b) illustrates $K$ electromagnetic-wave receivers, which are able to capture the electromagnetic waves in space and decode symbols.}
		\label{img:CAP}
		\vspace{-1em}
	\end{figure*}
As shown in Fig. \ref{img:CAP} (a), in this paper, we consider a \ac{cap-mimo} transmitter with surface ${\cal S}_{\rm T}$ of area $A_{\rm T}=\left|{\cal S}_{\rm T}\right|$. In the ideal case, \ac{cap-mimo} has a spatially-continuous antenna aperture, which is able to generate any current distribution on its continuous surface for wireless communications \cite{HuangHu'20,Sanguinetti'21,Decarli'21}. Consider that the \ac{cap-mimo} transmitter simultaneously serves $K$ electromagnetic-wave receivers in the downlink, as shown in Fig. \ref{img:CAP} (b). Let ${{\bf{x}}} \triangleq {\left[ {{x_{1}}, \cdots ,{x_{K}}} \right]^T} \in {{\mathbb C}^K}$ denote $K$ symbols transmitted to $K$ receivers, respectively. Without loss of generality, we assume that $\mathbb{E}_{\bf x}\left\{\mathbf{x}\mathbf{x}^{\rm H}\right\}=\mathbf{I}_{K}$. 

Following the well-known {\it time-harmonic assumption} \cite{Fred'08}, let ${\bf j}({\bf s})\in{\mathbb C}^{3}$ denote the time-independent current distribution at location ${\bf s}:=\left(s_x,s_y,s_z\right)\in{\mathbb R}^{3}$. Then, the symbols to be transmitted are modulated on $K$ different \ac{cap-mimo} patterns (i.e., current distributions), which aims to make these symbols orthogonal as much as possible so as to achieve high capacity \cite{HuangHu'20}. Assuming that \ac{cap-mimo} employs linear pattern combination, the information-carrying current distribution ${\bf{j}}({\bf{s}})$ on the \ac{cap-mimo} aperture can be modeled as 
	\begin{equation}\label{eqn:js_theta}
		{\bf{j}}({\bf{s}}) = \sum\limits_{k = 1}^K {{{\bm{\theta }}_k}\left( {\bf{s}} \right){x_k}},~~{\mathbf{s}\in{{\cal S}_{\rm T}}}, 
	\end{equation}
	where ${\bm{\theta }}_k\left( {\bf{s}} \right)\in{\mathbb C}^3$ is the pattern that carries symbol $x_k$.

To model the radiated information-carrying electromagnetic waves in space, we define ${\bf e}({\bf r})\in\mathbb{C}^3$ as the electric field at point ${\bf r}:=\left(r_x,r_y,r_z\right)\in\mathbb{R}^3$, which is induced by the current distribution ${\bf j}({\bf s})$ on the \ac{cap-mimo} aperture. According to Helmholtz wave equation \cite{Fred'08}, by introducing Green
function $\mathbf{G}(\mathbf{r}, \mathbf{s})\in{\mathbb C}^{3\times 3}$, the electric field ${\bf e}({\bf r})$ at point ${\bf r}$ of the receiver can be written as
\begin{equation}\label{eqn:er}
	\mathbf{e}(\mathbf{r})=\int_{{\cal S}_{\rm T}} \mathbf{G}(\mathbf{r}, \mathbf{s}) \mathbf{j}(\mathbf{s}) \mathrm{d} \mathbf{s},
\end{equation}
where Green function $\mathbf{G}(\mathbf{r}, \mathbf{s})$ is similar to the classical definition of wireless channels. Note that, $\mathbf{G}(\mathbf{r}, \mathbf{s})$ is determined by the specific transmission environment. For example, in ideal unbounded and homogeneous mediums, we have
\begin{equation}\label{eqn:free-space}
	\mathbf{G}(\mathbf{r}, \mathbf{s})=\frac{\mathrm{j} \kappa Z_{0}}{4 \pi} \frac{e^{\mathrm{j} \kappa\|\mathbf{r}-\mathbf{s}\|}}{\|\mathbf{r}-\mathbf{s}\|}\left(\mathbf{I}_3+\frac{\nabla_{\mathbf{r}} \nabla_{\mathbf{r}}^{\mathrm{H}}}{\kappa^{2}}\right),
\end{equation}
where $Z_{0}$ is the intrinsic impedance of spatial medium.


As shown in Fig. \ref{img:CAP} (b), we assume that all $K$ receivers are located in the far-field region, and each receiver is equipped with an ideal isotropic antenna with effective aperture area $A_{\rm R}=\frac{\lambda^2}{4\pi}$, which is much smaller than the transmitter aperture area $A_{\rm T}$. In this case, each receiver can be reasonably approximated by a point in space. Let ${\bf r}_k\in{\mathbb R}^3$ denote the 3-D location of the $k$-th electromagnetic-wave receiver. In the ideal case, the receiver $k$ is expected to ideally sense and capture the whole information of the electromagnetic waves reaching point ${\bf r}_k$. According to (\ref{eqn:js_theta}) and (\ref{eqn:er}), the electromagnetic wave captured by receiver $k$ can be expressed as
	\begin{equation}\label{eqn:signal_model}
		\begin{aligned}
			&{{\bf{y}}_k} = {{\bf{e}}_k} + {{\bf{n}}_k}\\
			&= \underbrace {{x_k}\!\!\int_{{{\cal S}_{\rm{T}}}} \!\!{{{\bf{G}}_k}({\bf{s}})} {\bm \theta _k}\left( {\bf{s}} \right){\rm{d}}{\bf{s}}}_{\text{Desired signal to user $k$}} + \!\!\!\! \underbrace {\sum\limits_{j = 1,j \ne k}^K\!\! {{x_j}\!\!\int_{{{\cal S}_{\rm{T}}}}\!\! {{{\bf{G}}_k}({\bf{s}}){\bm \theta _j}\left( {\bf{s}} \right)} {\rm{d}}{\bf{s}}} }_{\text{Interference from other receivers}} + \underbrace {{{\bf{n}}_k}}_{\text{Noise}},
		\end{aligned}
	\end{equation}
	where ${{\bf{e}}_k}:=\mathbf{e}(\mathbf{r}_k)$, ${{\bf{G}}_k}({\bf s}):=\mathbf{G}(\mathbf{r}_k, \mathbf{s})$, and ${{\bf{n}}_k}$ is the electromagnetic noise at receiver $k$, which is produced by all incoming electromagnetic waves that are not generated by the transmitter. Here we assume that, ${{\bf{n}}_k}$ is \ac{awgn} with zero mean and variance $\sigma^2{\bf I}_3$.

	\subsection{Sum-rate maximization problem formulation}
	Based on electromagnetic signal model (\ref{eqn:signal_model}), by calculating the mutual information \cite{Fred'08}, the \ac{cap-mimo} channel capacity, i.e., the sum-rate of $K$ receivers, can be written as
		\begin{equation}\label{eqn:SR}
			{R_{\rm sum}} = \sum\limits_{k=1}^K {{\log _2}\det \left| {{{\bf{I}}_3} + {{\bm{\alpha }}_k}{\bm{\alpha }}_k^{\rm{H}}{\bf{J}}_k^{-1}} \right|}, 
		\end{equation}
		where ${\bm{\alpha }}_k$ and ${\bf{J}}_k$ are given by 
		\begin{align}
			\label{eqn:alpha_J}
			{{\bm{\alpha }}_k} =& \int_{{{\cal S}_{\rm{T}}}} {{{\bf{G}}_k}({\bf{s}}){{\bm{\theta }}_k}\left( {\bf{s}} \right){\rm d}{\bf{s}}},\\
			{{\bf{J}}_k} = & \!\!\!\! \sum\limits_{j=1,j \ne k}^K { {\int_{\!{\cal S}_{\rm{T}}} \!\!\! {{{\bf{G}}_k}({\bf{s}}){{\bm{\theta }}_j}\left( {\bf{s}} \right){\rm d}{\bf{s}}} }{{\left( {\int_{\!{\cal S}_{\rm{T}}}\!\!\! {{{\bf{G}}_k}({\bf{s'}}){{\bm{\theta }}_j}\left( {\bf{s'}} \right){\rm d}{\bf{s'}}} } \right)}^{\!\!\!\rm{H}}}}  \!\!+\! {\sigma ^2}{{\bf{I}}_3}. \notag
		\end{align}
	
	In practical systems, we are interested in investigating the \ac{cap-mimo} channel capacity under a given power constraint. By integrating the radial component of the Poynting vector over a sphere with infinite-length radius, the physical transmit power of \ac{cap-mimo} based communication systems can be bounded by \cite{Fred'08}
		\begin{equation}\label{eqn:power_cons}
			\sum\limits_{k = 1}^K {\int_{{{\cal S}_{\rm{T}}}} {{{\left\| {{{\bm{\theta }}_k}\left( {\bf{s}} \right)} \right\|}^2}{\rm{d}}{\bf{s}}} }  \le {P_{\rm{T}}},
		\end{equation}
		where $P_{\rm T}$ can be viewed as the allowable maximum ``transmit power'' of \ac{cap-mimo}, which is implicitly associated with the physical energy and measured in [${\text A}^2$].

	By combing sum-rate (\ref{eqn:SR}) and transmit power constraint (\ref{eqn:power_cons}), the original problem of \ac{cap-mimo} sum-rate maximization subject to the power constraint can be formulated as
	\begin{subequations}\label{eqn:original_problem}
		\begin{align}
			\!\!\!\!{\cal P }_o:~~&\mathop{\max}\limits_{{{\bm \theta}}\left({\bf s}\right)}~~{R_{\rm sum}} = \sum\limits_{k=1}^K {{\log _2}\det \left| {{{\bf{I}}_3} + {{\bm{\alpha }}_k}{\bm{\alpha }}_k^{\rm{H}}{\bf{J}}_k^{-1}} \right|}, \label{eqn:objective} \\
			&~~{\rm s.t.}~~\sum\limits_{k = 1}^K {\int_{{{\cal S}_{\rm{T}}}} {{{\left\| {{{\bm{\theta }}_k}\left( {\bf{s}} \right)} \right\|}^2}{\rm{d}}{\bf{s}}} }  \le {P_{\rm{T}}}, \label{eqn:power}
		\end{align}
	\end{subequations}
	where ${\bm{\theta }}({\bf s})$ denotes the set of functions ${\bm{\theta }}_k({\bf s})$ for all $k\in\{1,\cdots,K\}$. Our goal is to maximize the sum-rate (\ref{eqn:objective}) by appropriately designing the continuous pattern functions ${\bm{\theta }}_k({\bf s})$ for all $k\in\{1,\cdots,K\}$. Note that, the sum-rate maximization problem ${\cal P }_o$ in (\ref{eqn:original_problem}) is difficult to solve. The reason is that, the coupled continuous functions ${\bm{\theta }}({\bf s})$ within the integral items exist in both optimization objective and constraint in (\ref{eqn:original_problem}), which is a non-convex density functional optimization problem. Such kind of continuous functional optimizations are common in the micro-wave area \cite{HuangHu'20}, where they are usually addressed by using the commercial electromagnetic simulation software such as high frequency structure simulator (HFSS), which leads to high time and space complexity. 

\section{Proposed Pattern-Division Multiplexing Technology}\label{sec:Alg}
	
	
\subsection{A general pattern-division multiplexing technique}\label{sec:method1}
To support \ac{cap-mimo} in general communication scenarios, different from the existing works which adopt the patterns generated by given special functions \cite{Decarli'21,Sanguinetti'21,HuangHu'20}, we propose the pattern-division multiplexing to flexibly design the patterns for \ac{cap-mimo} according to the real-time channel state information. Specifically, the proposed pattern-division multiplexing aims to make the information-carrying electromagnetic waves reaching different receivers orthogonal as much as possible. In this way, higher channel capacity is expected, which is similar to the space-division multiplexing in classical \ac{mimo} systems. 
	
To efficiently optimize the continuous pattern functions ${\bm{\theta }}_k({\bf s})$, an intuitive idea is to use series expansion to project these continuous functions onto an orthogonal space. Since Fourier series expansion has good generality, in this paper, we use Fourier bases to expand the continuous functions. For a clear illustration, we first introduce the following lemma.
	\begin{lemma}[Fourier series expansion]
		For any arbitrary continuous function $f(t)\in{\mathbb C}$ defined in $t\in[a,b]\in{\mathbb R}$, if $f(t)$ is absolutely integrable, $f(t)$ can be equivalently rewritten as
		\begin{equation}
			f\left( t \right) = \sum\limits_{n =  - \infty }^\infty  {{F_n}{{\Psi _n}\left( t \right)}}  ,
		\end{equation}
		where the projection length $F_n\in{\mathbb C}$ and the Fourier basis ${\Psi _n}\left( t \right)\in{\mathbb C}$ are respectively given by
		\begin{subequations}
			\begin{align}
				{F_n} =& \frac{1}{{\sqrt {b - a} }}\int_a^b {f\left( t \right){e^{\frac{-{{\rm{j}}2\pi n}}{{b - a}}\left( {t - \frac{{b - a}}{2}} \right)}}} {\rm{d}}t,\\
				{\Psi _n}\left( t \right) =& \frac{1}{{\sqrt {b - a} }}{e^{\frac{{{\rm{j}}2\pi n}}{{b - a}}\left( {t - \frac{{b - a}}{2}} \right)}}.
			\end{align}
		\end{subequations}
	\end{lemma}
	
	Then, by extending the above {\it Lemma 1} to 3-D scenario, the continuous pattern functions ${\bm \theta}_k({\bf s})$ to be designed can be equivalently rewritten as
	\begin{equation}\label{eqn:theta_f}
		{\bm \theta}_k({\bf s})=\sum\limits_{\bf{n}}^\infty  {{{\bf{w}}_{k,{\bf{n}}}}{\Psi _{\bf{n}}}\left( {\bf{s}} \right)},~~{\bf s} \in {\cal S}_{\rm T},
	\end{equation}
	where ${\bf{w}}_{k,{\bf{n}}}\in{\mathbb C}^3$ is the projection length of pattern ${\bm \theta}_k({\bf s})$ on the Fourier basis ${\Psi _{\bf{n}}}\left( {\bf{s}} \right)\in{\mathbb C}$, and here we define ${\bf{n}} := {\left({n_x},{n_y},{n_z}\right)}$ to distinguish different expansion items and $\sum\nolimits_{\bf{n}}^\infty {} :=  \sum\nolimits_{{n_x} =  - \infty}^\infty {\sum\nolimits_{{n_y} =  - \infty}^\infty {\sum\nolimits_{{n_z} =  - \infty}^\infty {} } }$ for expression simplification. In 3-D scenarios, Fourier basis ${\Psi _{\bf{n}}}\left( {\bf{s}} \right)$ becomes
	\begin{equation}
	{\Psi _{\bf{n}}}\!\left( {\bf{s}} \right) = \frac{1}{\sqrt{A_{\rm T}}}{e^{ {\rm j}2\pi \left( {\frac{{{n_x}}}{{{L_x}}}{{s_x}}  + \frac{{{n_y}}}{{{L_y}}}{{s_y}} + \frac{{{n_z}}}{{{L_z}}} {{s_z}} } \right)}}
	\end{equation}
	wherein $L_x$, $L_y$, and $L_z$ denote the maximum projection lengths of \ac{cap-mimo} aperture area $A_{\rm T}$ on the $x$-, $y$-, and $z$-axis of 3-D coordinate system, respectively. Relying on this transformation, we obtain the following two corollaries.
	\begin{corollary}[Continuous-discrete transformation for electromagnetic waves] 
		By adopting Fourier series expansion, the coupled Green function ${\bf{G}}_k({\bf s})$ and pattern function ${\bm{\theta }}_j({\bf s})$ in integral $\int_{{{\cal S}_{\rm T}}}{\rm d}{\bf s}$ (i.e., the electromagnetic wave), can be equivalently rewritten as
		\begin{equation}\label{eqn:Green_Four}
			\begin{aligned}
				\int_{{{\cal S}_{\rm{T}}}} {{{\bf{G}}_k}({\bf{s}}){{\bm{\theta }}_j}\left( {\bf{s}} \right){\rm d}{\bf{s}}} = \sum\limits_{\bf{n}}^\infty  {{{\bf{\Omega }}_{k,{\bf{n}}}}{{\bf{w}}_{j,{\bf{n}}}}},
			\end{aligned}
		\end{equation}
		where
		\begin{equation}
			\begin{aligned}
				{{\bf{\Omega }}_{k,{\bf{n}}}} = \int_{{{\cal S}_{\rm{T}}}} {{{\bf{G}}_k}({\bf{s}}){\Psi _{\bf{n}}}\left( {\bf{s}} \right){\rm{d}}{\bf{s}}},
			\end{aligned}
		\end{equation}
		and ${{{\bf{\Omega }}_{k,{\bf{n}}}}}\in{\mathbb C}^{3\times 3}$ is exactly the Fourier transform of Green function ${\bf{G}}_k({\bf s})$ on the \ac{cap-mimo} aperture ${\cal S}_{\rm T}$ at the spatial frequency of $(\frac{{n_x}}{L_x},\frac{{n_y}}{L_y},\frac{{n_z}}{L_z})$. 
	\end{corollary}

	\begin{corollary}[Continuous-discrete transformation for power constraint] 
		According to Parseval's theorem, the square $L^2$ norm of current density ${\bm{\theta }}_k({\bf s})$ in integral expressions, i.e., the transmit power, can be equivalently rewritten as
		\begin{equation}\label{eqn:approx_theta}
			\begin{aligned}
				\int_{{{\cal S}_{\rm{T}}}} {{{\left\| {{{\bm{\theta }}_k}\left( {\bf{s}} \right)} \right\|}^2}{\rm{d}}{\bf{s}}}  = \sum\limits_{\bf{n}}^\infty  {{{\left\| {{{\bf{w}}_{k,{\bf{n}}}}} \right\|}^2}}.
			\end{aligned}
		\end{equation}
	\end{corollary}
	
	Exploiting {\it Corollary 1} and {\it Corollary 2}, the design of the continuous pattern functions ${\bm \theta}_k({\bf s})$ becomes the design of the projection lengths ${\bf{w}}_{k,{\bf{n}}}$, and thus the density functional optimization can be equivalently transformed as a common digital signal processing problem. However, since the number of expansion items is infinite, the projection lengths ${\bf{w}}_{k,{\bf{n}}}$ are still hard to be optimized. Fortunately, thanks to the mathematical structure of Green function ${\bf{G}}_k({\bf s})$, we notice that, the power of ${\bf{G}}_k({\bf s})$ in Fourier space is mainly distributed in the low-frequency band, while that in the high-frequency band is very low (see \cite[Fig. 4]{Sanguinetti'21}). In other words, when the number of expansion items $\bf n$ is large enough, the value of ${\bf{\Omega }}_{k,{\bf{n}}}$ tends to be negligible. This fact inspires us to approximate the original continuous function with finite Fourier expansion items, thus we obtain the following proposition.
	
	\begin{proposition}[Finite-item approximation of Fourier series] 
	Employing a truncation operation on (\ref{eqn:Green_Four}) and (\ref{eqn:approx_theta}),	
	the electric field and transmit power can be approximated by
		\begin{subequations}\label{eqn:truncation}
			\begin{align}
				\int_{{{\cal S}_{\rm{T}}}} \!\!{{{\bf{G}}_k}({\bf{s}}){{\bm{\theta }}_j}\left( {\bf{s}} \right){\rm d}{\bf{s}}} &\approx \sum\limits_{\bf{n}}^{\bf N}  {{{\bf{\Omega }}_{k,{\bf{n}}}}{{\bf{w}}_{j,{\bf{n}}}}}, \\
				\int_{{{\cal S}_{\rm{T}}}} {{{\left\| {{{\bm{\theta }}_k}\left( {\bf{s}} \right)} \right\|}^2}{\rm{d}}{\bf{s}}}  &\approx \sum\limits_{\bf{n}}^{\bf N}  {{{\left\| {{{\bf{w}}_{k,{\bf{n}}}}} \right\|}^2}},
			\end{align}
		\end{subequations}
		where we have ${\bf N}:=\left(N_x,N_y,N_z\right)$ with $N_x$, $N_y$, and $N_z$ being the numbers of reserved items, and we have $\sum\nolimits_{\bf{n}}^{\bf{N}} :  = \sum\nolimits_{{n_x} = \left\lfloor { - {N_x}/2} \right\rfloor }^{\left\lfloor {{N_x}/2} \right\rfloor } {\sum\nolimits_{{n_y} = \left\lfloor { - {N_y}/2} \right\rfloor }^{\left\lfloor {{N_y}/2} \right\rfloor } {\sum\nolimits_{{n_z} = \left\lfloor { - {N_z}/2} \right\rfloor }^{\left\lfloor {{N_z}/2} \right\rfloor } {} } } $ for clarity.
	\end{proposition}

	\subsection{Proposed pattern design scheme for sum-rate maximization}\label{sec:method2}
	Based on the proposed pattern-division multiplexing technique introduced above, in this subsection, we propose a pattern design scheme to solve the sum-rate maximization problem ${\cal P }_o$ in (\ref{eqn:original_problem}). Firstly, to decouple pattern functions ${\bm{\theta }}_k({\bf s})$, by adopting an equivalent transformation \cite[\it Theorem 1]{Qingjiang'11} for ${\cal P }_o$ in (\ref{eqn:original_problem}), we obtain the following lemma.
	\begin{algorithm}[!b] 
		\caption{Proposed pattern design scheme.} 
		\label{alg:1} 
		\begin{algorithmic}[1] 
			\REQUIRE ~~ 
			Green functions ${{\bm{G}}_{k}\left({\bf s}\right)}$ for all $k\in\{1,\cdots,K\}$.
			\ENSURE ~~ 
			Optimized $R_{{\rm sum}}$, $\bm{\psi}$, and $\bm{\theta}\left({\bf s}\right)$.
			\STATE Initialize $\bm{\psi}$ and $\bm{\theta}\left({\bf s}\right)$;
			\WHILE {No convergence of $R_{{\rm sum}}$}
			\STATE Update ${\bm \rho}$ by (\ref{eqn:rho_update});
			\STATE Update $\bm{\psi}$ by (\ref{eqn:psi_update});
			\STATE Update ${\bf w}$ by (\ref{eqn:w_update}) and (\ref{eqn:zeta});
			\STATE Update $\bm{\theta}\left({\bf s}\right)$ by (\ref{eqn:w_2_theta});
			\ENDWHILE	
			\RETURN Optimized $R^{\rm opt}_{{\rm sum}}$, $\bm{\psi}^{\rm opt}$, and $\bm{\theta}^{\rm opt}\left({\bf s}\right)$. 
		\end{algorithmic}
	\end{algorithm}
	
	\begin{lemma}[Equivalent problem for sum-rate maximization]
		By introducing an auxiliary variable ${\bm{\rho }} = {\left[ {{\rho _k}, \cdots ,{\rho _K}} \right]^{\rm{T}}}\in{\mathbb R}^K_{+}$ and the combining vectors ${\bm{\psi }} = \left[ {{\bm\psi _1}, \cdots ,{\bm\psi _K}} \right]\in{\mathbb C}^{3\times K}$, the original sum-rate maximization problem ${\cal P }_o$ in (\ref{eqn:original_problem}) can be equivalently reformulated as 
		\begin{subequations}\label{eqn:Rsum'}
			\begin{align}
				\!\!\!\!{\cal P }_1:&\mathop{\max}\limits_{{\bm{\rho }}, {\bm \psi}, {{\bm \theta}}\left({\bf s}\right)}{R_{{\rm{sum}}}'} \!=\! \sum\limits_{k = 1}^K {{{\log }_2}{\rho _k}}  \!-\! \frac{1}{{\ln 2}}\sum\limits_{k = 1}^K {{\rho _k}{E_k}}  \!+\! \frac{K}{{\ln 2}},\\
				&~~{\rm s.t.}~~\sum\limits_{k = 1}^K {\int_{{{\cal S}_{\rm{T}}}} {{{\left\| {{{\bm{\theta }}_k}\left( {\bf{s}} \right)} \right\|}^2}{\rm{d}}{\bf{s}}} }  \le {P_{\rm{T}}},
			\end{align}
		\end{subequations}
		where $E_k:={{\mathbb E}_{{\bf{x}},{\bf{n}}}}\left\{ {\left| {{{\hat x}_k} - {x_k}} \right|}^2 \right\}$ is the mean-square error (MSE) of the decoded symbol ${\hat x}_k={\bm \psi}_k^{\rm H}{\bf y}_k$, defined as
		\begin{equation}
			\begin{aligned}
				{E_k} =& {\left| {1 - \int_{{S_{\rm{T}}}} {{\bm{\psi }}_k^{\rm{H}}{{\bf{G}}_k}({\bf{s}})} {{\bm\theta}_k}\left( {\bf{s}} \right){\rm{d}}{\bf{s}}} \right|^2} + \\ & \sum\limits_{j = 1,j \ne k}^K {{{\left| {\int_{{S_{\rm{T}}}} {{\bm{\psi }}_k^{\rm{H}}{{\bf{G}}_k}({\bf{s}}){{\bm{\theta }}_j}\left( {\bf{s}} \right)} {\rm{d}}{\bf{s}}} \right|}^2}}  + {\sigma ^2}{\left\| {{{\bm{\psi }}_k}} \right\|^2}.
			\end{aligned}
		\end{equation}
	\end{lemma}
	
	To solve the equivalent problem ${\cal P }_1$ in (\ref{eqn:Rsum'}), a scheme of pattern design can be established by optimizing variables $\bm \rho$, combiners $\bm \psi$, and continuous functions ${\bm \theta}({\bf s})$ alternatively until the convergence of sum-rate $R_{\rm sum}$. For clarity, we summarize the whole process of this pattern design scheme in {\bf Algorithm 1}, where the update steps of  $\bm \rho$, $\bm \psi$, and ${\bm \theta}({\bf s})$ will be introduced in the following three parts, respectively.

	\subsubsection{Fix $\bm \psi$ and ${\bm \theta}({\bf s})$, then optimize ${\bm \rho}$}
	While fixing the combiners $\bm \psi$ and patterns ${\bm \theta}({\bf s})$, the optimal solution to ${\bm \rho}$ can be obtained by setting $\frac{{\partial {R'_{\rm sum}}}}{{\partial {\rho _k}}}$ to zero, given by
	\begin{equation}\label{eqn:rho_update}
		{\rho _k^{\rm opt}} = E_k^{ - 1},~~~k\in\{1,\cdots,K\}.
	\end{equation}
	\subsubsection{Fix ${\bm \rho}$ and ${\bm \theta}({\bf s})$, then optimize $\bm \psi$}
	While fixing ${\bm \rho}$ and patterns ${\bm \theta}({\bf s})$, by defining 
\begin{align*}
	&{{\bf{A}}_k} \!=\! \rho_k\!\!\sum\limits_{j = 1}^K \!\!{\int_{\!{{\cal S}_{\rm{T}}}}\!\! {{{\bf{G}}_k}({\bf{s}}){{\bm{\theta }}_j}\left( {\bf{s}} \right)} {\rm{d}}{\bf{s}}{{\left( {\int_{\!{{\cal S}_{\rm{T}}}}\!\!\!\! {{{\bf{G}}_k}({{\bf{s}}^\prime }){{\bm{\theta }}_j}\left( {{{\bf{s}}^\prime }} \right)} {\rm{d}}{{\bf{s}}^\prime }} \right)}^{\!\!\rm{H}}}}  \!\!\!\!+\! {\rho_k}{\sigma ^2}{{\bf{I}}_3}, \\
	&{{\bm{\alpha }}_k} \!=\! \rho_k\int_{{{\cal S}_{\rm{T}}}} {{{\bf{G}}_k}({\bf{s}}){{\bm{\theta }}_k}\left( {\bf{s}} \right){\rm d}{\bf{s}}}.
\end{align*}
and solving $\frac{{\partial {R'_{\rm sum}}}}{{\partial {{\bm{\psi }}_k}}}=0$, the optimal solution to ${\bm{\psi }}_k$ can be easily calculated as
	\begin{align}\label{eqn:psi_update}
		{{\bm{\psi }}_k^{\rm opt}} = {\bf{A}}_k^{ - 1}{{\bm{\alpha }}_k},~~~k\in\{1,\cdots,K\}.
	\end{align}
	\subsubsection{Fix ${\bm \rho}$ and $\bm \psi$, then optimize ${\bm \theta}({\bf s})$}
	Given fixed ${\bm \rho}$ and combiners $\bm \psi$, after removing the unrelated components, the subproblem of optimizing the continuous pattern functions ${\bm \theta}({\bf s})$ can be reformulated as
	\begin{subequations}\label{eqn:problem_p3}
		\begin{align}
			\!\!\!\!{\cal P }_3:~~&\mathop{\max}\limits_{{\bm \theta}\left({\bf s}\right)}~~ \sum\limits_{k = 1}^K {\rho_kg_k\left({\bm \theta}\left({\bf s}\right)\right)},\\
			&~~{\rm s.t.}~~\sum\limits_{k = 1}^K {\int_{{{\cal S}_{\rm{T}}}} {{{\left\| {{{\bm{\theta }}_k}\left( {\bf{s}} \right)} \right\|}^2}{\rm{d}}{\bf{s}}} }  \le {P_{\rm{T}}},
		\end{align}
	\end{subequations}
	where function ${g_k\left({\bm \theta}\left({\bf s}\right)\right)}$ is defined as
	\begin{equation}
		\begin{aligned}
			{g_k}\left( {{\bm{\theta }}\left( {\bf{s}} \right)} \right) =& \sum\limits_{j = 1}^K {{{\left| {\int_{{{\cal S}_{\rm{T}}}} {{\bm{\psi }}_k^{\rm{H}}{{\bf{G}}_k}({\bf{s}}){{\bm{\theta }}_j}\left( {\bf{s}} \right)} {\rm{d}}{\bf{s}}} \right|}^2}}  - \\ & ~~~~~~2\,{\mathop{\Re}\nolimits} \left\{ {\int_{{{\cal S}_{\rm{T}}}} {{\bm{\psi }}_k^{\rm{H}}{{\bf{G}}_k}({\bf{s}})} {{\bm{\theta }}_k}\left( {\bf{s}} \right){\rm{d}}{\bf{s}}} \right\}.
		\end{aligned}
	\end{equation}
	
	Next, to address the challenging issue of density functional optimization as shown in problem ${\cal P }_3$ in (\ref{eqn:problem_p3}), we apply the continuous-discrete transformations in {\it Corollary 1} and {\it Corollary 2}, and {\it Proposition 1} to reformulate problem ${\cal P }_3$ in (\ref{eqn:problem_p3}), and the reformulated problem can be written as	
	\begin{subequations}\label{eqn:problem_p4}
		\begin{align}
			\!\!\!\!{\cal P }_4:~~&\mathop{\max}\limits_{\bf w}~~ \sum\limits_{k = 1}^K {\rho_k{{\hat g}_k}\left( {\bf{w}} \right)},\\
			&~~{\rm s.t.}~~\sum\limits_{k = 1}^K {\sum\limits_{\bf{n}}^{\bf N} {{{\left\| {{{\bf{w}}_{k,{\bf{n}}}}} \right\|}^2}} }  \le {P_{\rm{T}}},
		\end{align}
	\end{subequations}
	where we have defined ${\bf{w}}$ as the set of ${\bf{w}}_{k,{\bf n}}$ and
	\begin{align}
		{{\hat g}_k}\left( {\bf{w}} \right) \!=\! \sum\limits_{j = 1}^K {{{\left| {\sum\limits_{\bf{n}}^{\bf N} {{\bf{h}}_{k,{\bf{n}}}^{\rm{H}}{{\bf{w}}_{j,{\bf{n}}}}} } \right|}^2}}  - 2{\mathop{\Re}\nolimits} \left\{ {\sum\limits_{\bf{n}}^{\bf N} {{\bf{h}}_{k,{\bf{n}}}^{\rm{H}}{{\bf{w}}_{k,{\bf{n}}}}} } \right\},
	\end{align}
	in which ${{\bf{h}}_{k,{\bf{n}}}} := {\bf{\Omega }}_{k,{\bf{n}}}^{\rm{H}}{{\bm{\psi }}_k}$.
	\par
	To further simplify the expression, we define ${\bf h}_k$  and ${\bf w}_k$ as the vectorized sets of ${\bf h}_{k,{\bf n}}$ and ${\bf w}_{k,{\bf n}}$ for all ${\bf n}=[n_x,n_y,n_z]\in\left\{\{-N_x,\cdots,N_x\},\{-N_y,\cdots,N_y\},\{-N_z,\cdots,N_z\}\right\}$. Thus problem ${\cal P }_4$ in (\ref{eqn:problem_p4}) can be equivalently reorganized as
	\begin{subequations}\label{eqn:problem_p5}
		\begin{align}
			\!\!\!\!{\cal P }_5:~~&\mathop{\max}\limits_{{\bf w}}~~ \sum\limits_{k = 1}^K \rho_k {\left( {\sum\limits_{j = 1}^K {{{\left| {{\bf{h}}_k^{\rm{H}}{{\bf{w}}_j}} \right|}^2}}  \!- 2{\mathop{\Re}\nolimits} \left\{ {{\bf{h}}_k^{\rm{H}}{{\bf{w}}_k}} \right\}} \right)} ,\\
			&~~{\rm s.t.}~~\sum\limits_{k = 1}^K {{{\left\| {{{\bf{w}}_k}} \right\|}^2}}  \le {P_{\rm{T}}}, \label{eqn:power_cons2}
		\end{align}
	\end{subequations}
	which is a standard \ac{qcqp}. By adopting Lagrange multiplier method, the optimal solution to problem ${\cal P }_5$ in (\ref{eqn:problem_p5}) is given by
	\begin{equation}\label{eqn:w_update}
		\begin{aligned}
			{{\bf{w}}_k^{\rm opt}} = {\rho _k}{\left( {{\rho _k}\sum\limits_{j = 1}^K {{{\bf{h}}_j}{\bf{h}}_j^{\rm{H}} + \zeta {{\bf{I}}_{3{N_F}}}} } \right)^{\!\!- 1}}{{\bf{h}}_k},
		\end{aligned}
	\end{equation}
	wherein $N_F:=(N_x+1)(N_y+1)(N_z+1)$ is the total number of Fourier expansion items. Note that $\zeta$ is the Lagrange multiplier, which should be chosen such that the complementarity slackness condition of power constraint (\ref{eqn:power_cons2}) is satisfied, i.e.,
	\begin{equation}\label{eqn:zeta}
		{\zeta ^{\rm opt}} = \min \left\{ {{\zeta} \ge 0:\sum\limits_{k = 1}^K {{{\left\| {{{\bf{w}}_k}} \right\|}^2}}  \le {P_{\rm{T}}}} \right\}.
	\end{equation}
	One-dimensional binary search can be an efficient way to solve (\ref{eqn:zeta}) and obtain the optimal multiplier $\zeta^{\rm opt}$.
	
	After calculating the optimal projection lengths ${{\bf{w}}_k^{\rm opt}}$, according to (\ref{eqn:theta_f}), the final solution to the \ac{cap-mimo} patterns can be obtained by
	\begin{equation}\label{eqn:w_2_theta}
		{\bm \theta}_k^{\rm opt}({\bf s})=\sum\limits_{\bf{n}}^{\bf N}  {{{\bf{w}}^{\rm opt}_{k,{\bf{n}}}}{\Psi _{\bf{n}}}\left( {\bf{s}} \right)},~~{\bf s} \in {\cal S}_{\rm T},
	\end{equation}
	which completes the proposed pattern design scheme.
	
	\section{Simulation Results}\label{sec:NSR}
	
	\subsection{Simulation setup}
	
	We consider a 3-D scenario for simulation, where one \ac{cap-mimo} transmitter simultaneously serves $K=8$ receivers. We assume that the \ac{cap-mimo} transmitter is deployed on the $xy$-plane with its center located at $\left(0,0,0\right)$, i.e.,
	\begin{equation}
		{{\cal S}_T}: = \left\{ { {\left( {{s_x},{s_y},{s_z}} \right)} {\Big |}\left| {{s_x}} \right| \le \frac{{{L_x}}}{2},\left| {{s_y}} \right| \le \frac{{{L_y}}}{2},{s_z} = 0} \right\},
	\end{equation}
	where the \ac{cap-mimo} aperture has a square shape with the area of $A_{\rm T}=0.25\,{\rm m}^2$, i.e., $L_x=L_y=0.5\,{\rm m}$. All receivers are located within a square region, where four electromagnetic-wave receivers are located at $\left(\pm 1\,{\rm m}, \pm 1\, {\rm m}, 30\, {\rm m}\right)$, and the other four receivers are located at $\left(\pm 5\,{\rm m}, \pm 5\,{\rm m}, 30\,{\rm m}\right)$, respectively. The frequency is set to $f=2.4$ $ {\rm GHz}$, and the intrinsic impedance is set to $Z_{0}=376.73$ $\Omega$ \cite{Fred'08}. Unless specially specified, the maximum transmit power is set to $P_{\rm T} = 10^{-4}$ ${\rm A}^2$, and the noise power is set to $\sigma^2 = 5.6\times10^{-3}$ ${\rm V}^2/{\rm m}^2$. The Green functions ${\bf G}_k\left({\bf s}\right)$ are generated by the free-space model (\ref{eqn:free-space}). The sampling number of the integral operation $\int_{{{\cal S}_{\rm T}}}{\rm d}{\bf s}$ is set to $I_s=1024$. The numbers of the reserved Fourier expansion items are set to $N_x=5$, $N_y=5$, and $N_z=0$, respectively, i.e., $N_F=36$. All pattern functions ${\bm \theta}_k({\bf s})$ and receiver combiners ${\bm \psi}_k$ are randomly initialized. 
	
	As for the benchmark scheme, we consider the existing wavenumber-division multiplexing recently proposed in \cite{Sanguinetti'21} for comparison, of which the key idea is to directly adopt multiple orthogonal Fourier basis functions to generate the patterns for different receivers respectively, as discussed in Section \ref{sec:intro}. To compare the interference elimination ability of different schemes, we consider the interference-free sum-rate as the upper bound for comparison, which can be realized by assuming that all inter-receiver interference can be ideally eliminated and then employing the proposed pattern design.
	\subsection{Sum-rate against aperture area $A_{\rm T}$}
	\begin{figure}[!t]
		\centering
		\includegraphics[width=3.6in]{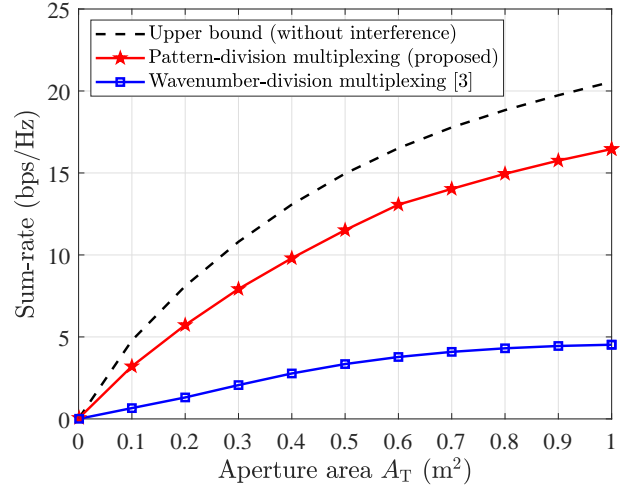}
		\caption{Sum-rate against the aperture area $A_{\rm T}$ of \ac{cap-mimo}.}
		\label{img:simulation_size}
		\vspace{-1em}
	\end{figure}
	We plot the sum-rate against the aperture area $A_{\rm T}$ in Fig. \ref{img:simulation_size}, where the aperture shape of \ac{cap-mimo} always remains a square, i.e., $L_x=L_y$. From this figure, we can observe that, for all schemes, the sum-rate quickly improves as the aperture area $A_{\rm T}$ increases. However, compared with the existing wavenumber-division multiplexing, the proposed pattern-division multiplexing achieves a higher sum-rate. For example, when $A_{\rm T}=1\,{\rm m}^2$, the sum-rate achieved by the proposed scheme is 16.46 bps/Hz, which is about 260\% higher than 4.53 bps/Hz achieved by the existing wavenumber-division multiplexing. The reason is that, the wavenumber-division multiplexing directly uses $K$ different orthogonal Fourier bases to generate $K$ different patterns. Despite this scheme can make the symbols orthogonal at the transmitter, their orthogonality cannot be guaranteed at the receivers. After passing through the electromagnetic channels, the orthogonality among different symbols may be destroyed, which will result in high inter-receiver interference or even make the radiated electromagnetic wave unable to reach the receivers accurately. By contrast, the proposed pattern-division multiplexing jointly designs the patterns for $K$ receivers according to the specific channel state information. During the process of pattern design, the proposed scheme actually makes a trade-off between the amplification of desired signals and the elimination of inter-receiver interference, and it can also achieve a power allocation among different users. In this way, the electromagnetic waves carrying symbols can be focused on the receivers accurately with stronger orthogonality, and thus higher sum-rate can be achieved.
	
	\subsection{Patterns ${\bm \theta}({\bf s})$ of \ac{cap-mimo}}
	\begin{figure}[!t]
		\setlength{\abovecaptionskip}{-0.0cm}
		\setlength{\belowcaptionskip}{-0.0cm}
		\centering
		\subfigcapskip -1em
		\subfigure[$k=1$]{
			\includegraphics[width=1.6in]{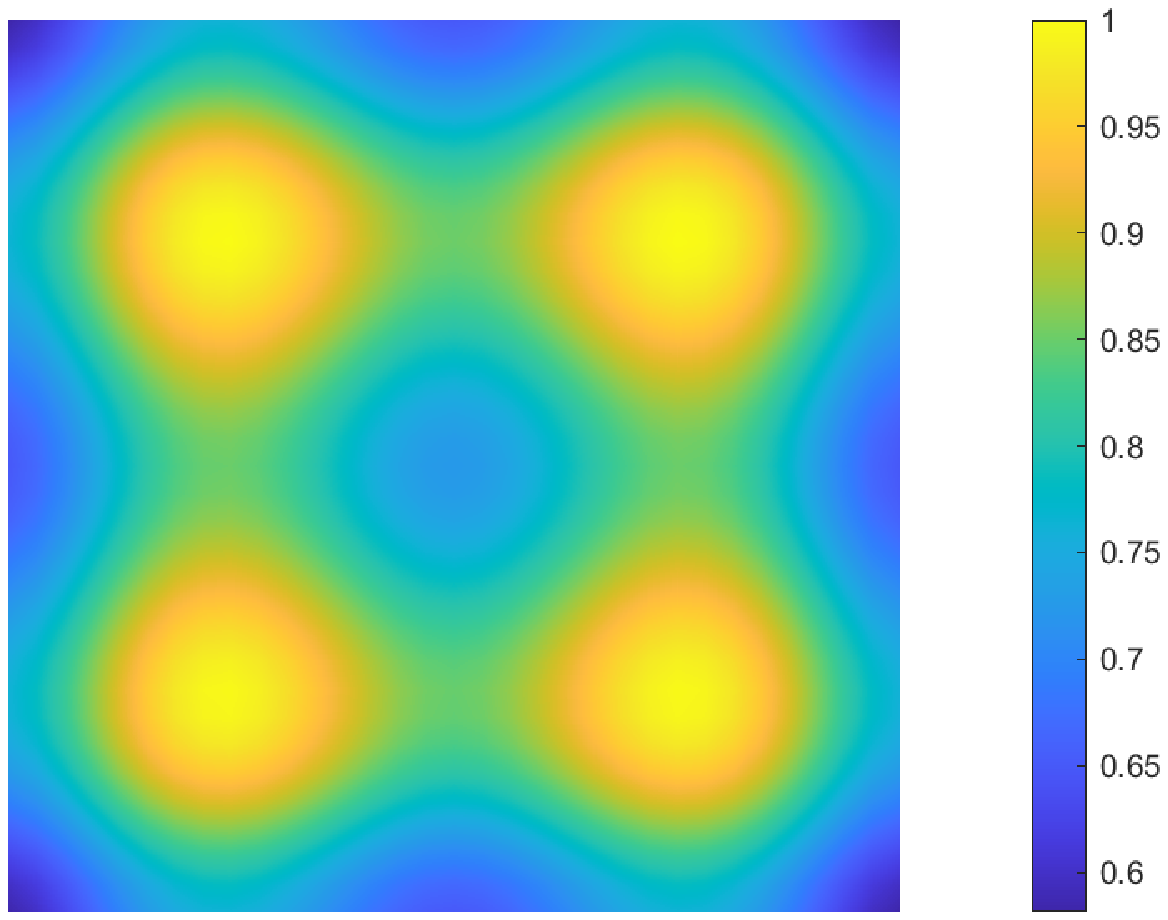}
		}\hspace{-2em}
		\subfigure[$k=2$]{
			\includegraphics[width=1.6in]{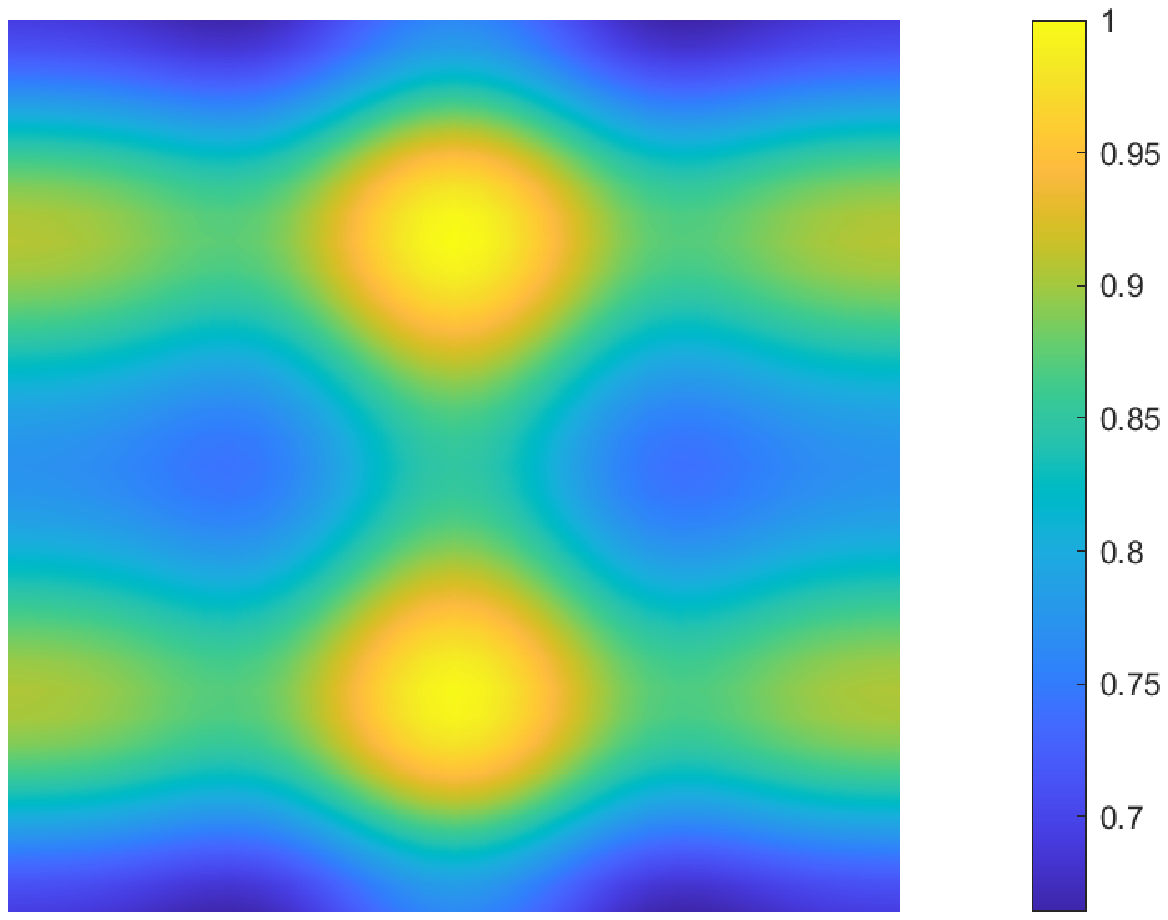}
		}\vspace{-1em}\\
		\subfigure[$k=3$]{
			\includegraphics[width=1.6in]{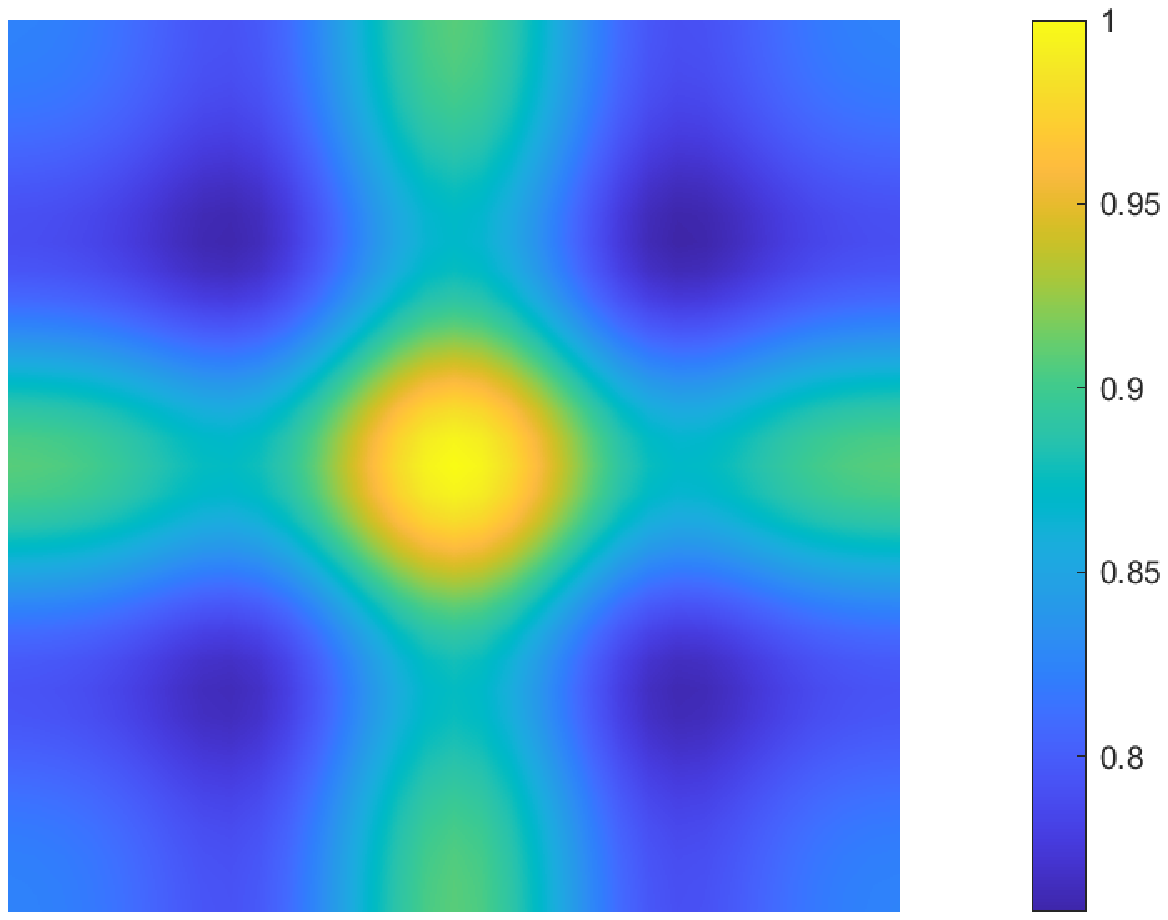}
		}\hspace{-2em}
		\subfigure[$k=4$]{
			\includegraphics[width=1.6in]{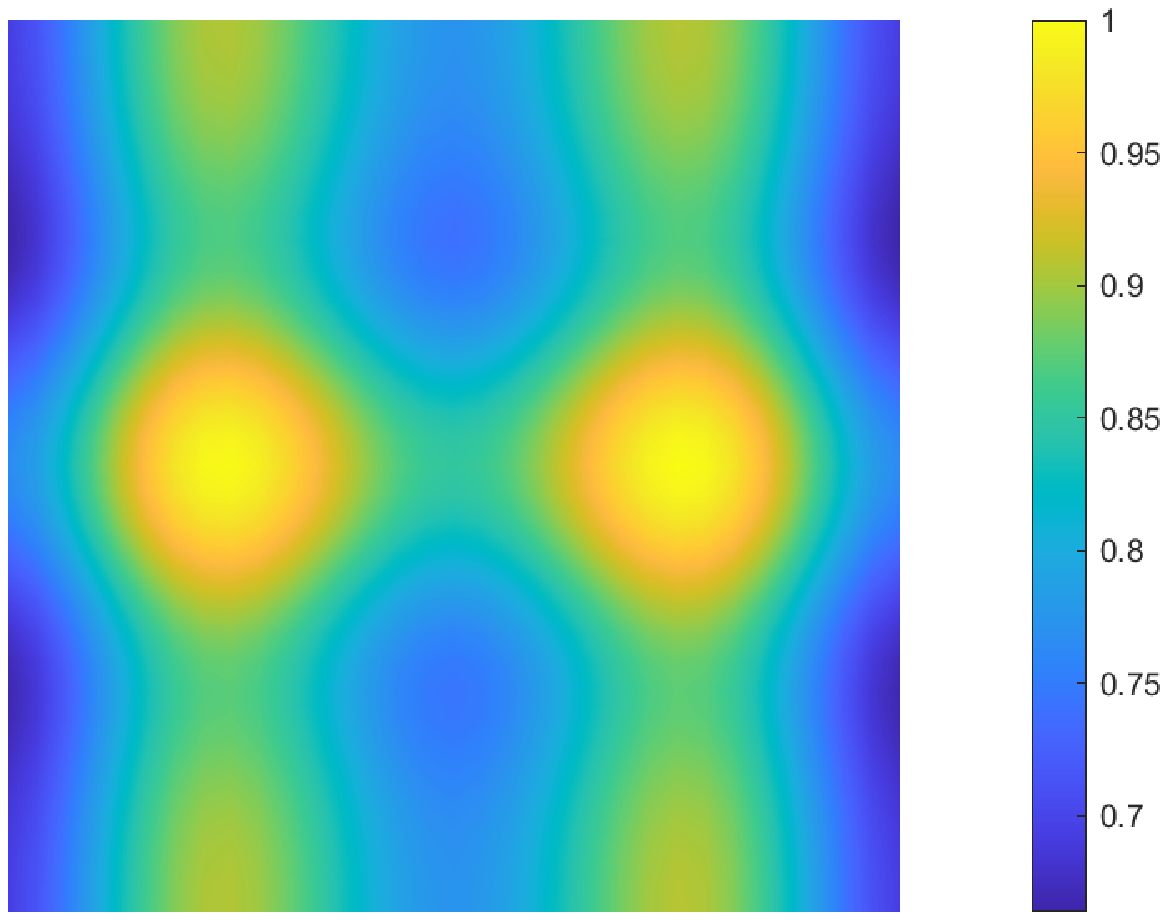}
		}
		\vspace{0em}
		\caption{
			The normalized amplitude of the x-component of the optimized pattern ${\bm \theta}_k({\bf s})$.
		}
		\label{img:simulation_amplitude}
		\vspace{-1em}
	\end{figure}
\begin{figure}[!t]
	\setlength{\abovecaptionskip}{-0.0cm}
	\setlength{\belowcaptionskip}{-0.0cm}
	\centering
	\subfigcapskip -1em
	\subfigure[$k=1$]{
		\includegraphics[width=1.6in]{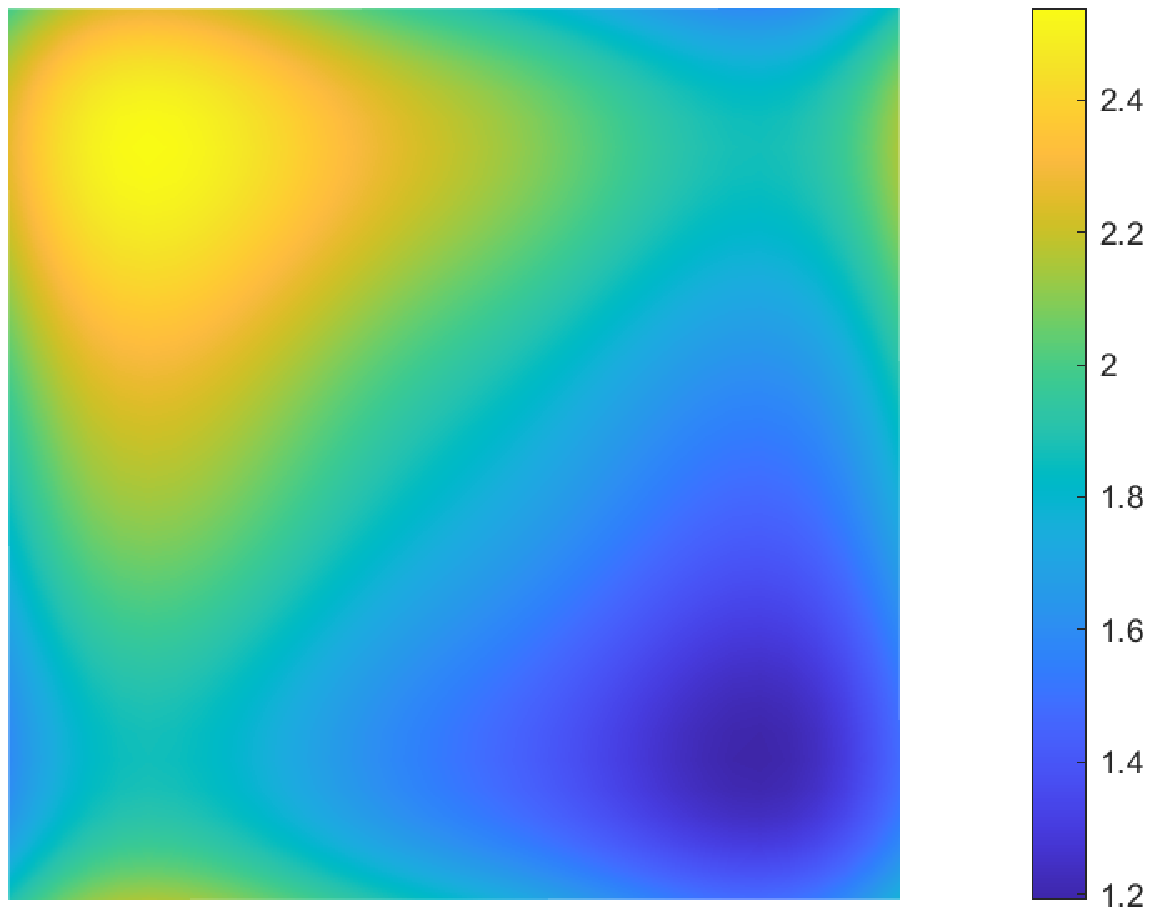}
	}\hspace{-2em}
	\subfigure[$k=2$]{
		\includegraphics[width=1.6in]{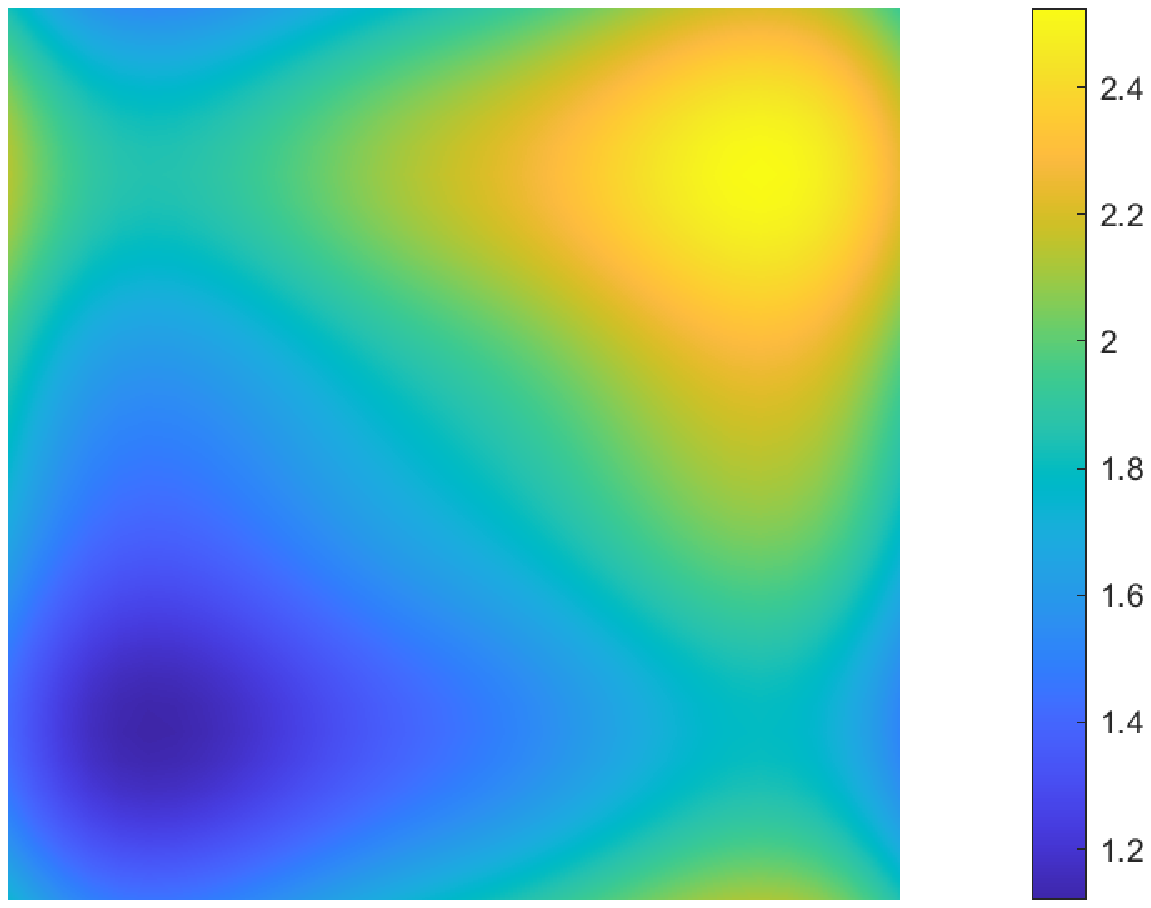}
	}\vspace{-1em}\\
	\subfigure[$k=3$]{
		\includegraphics[width=1.6in]{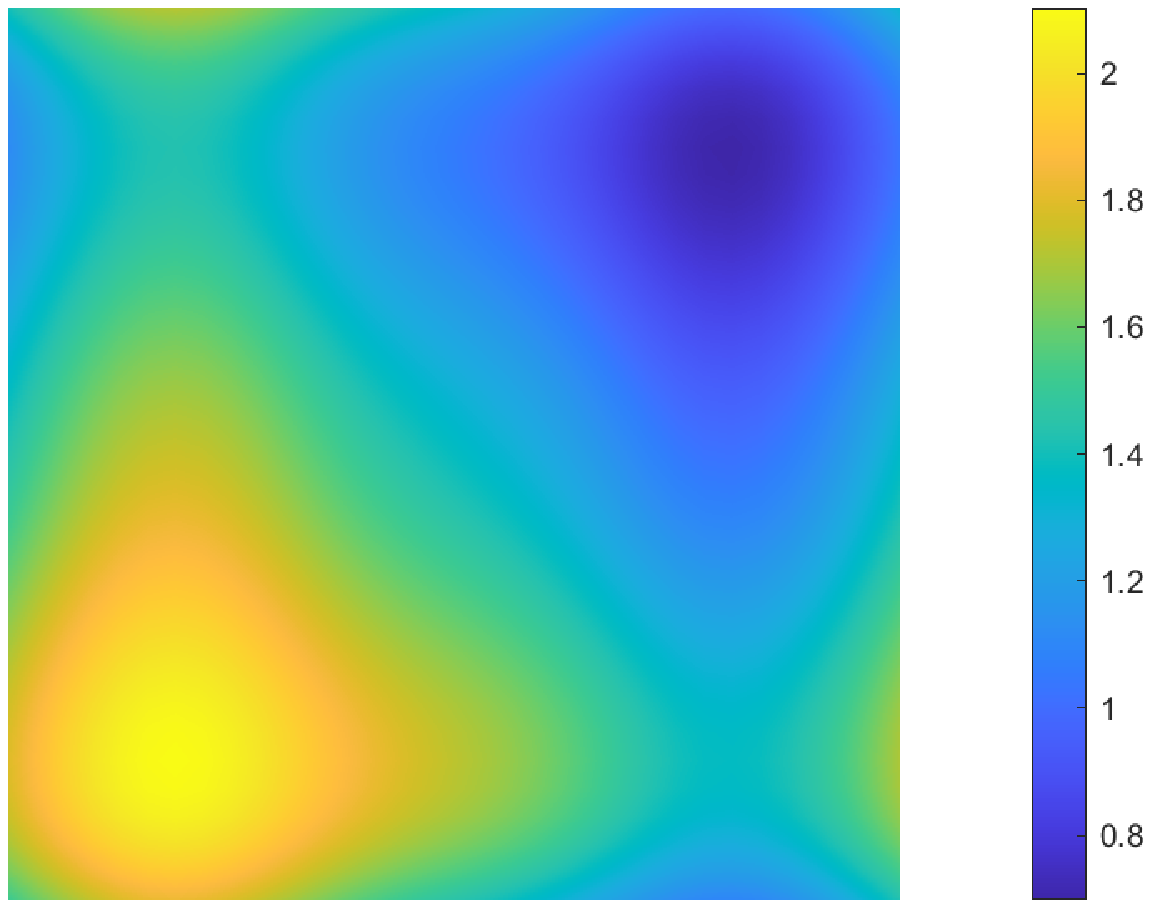}
	}\hspace{-2em}
	\subfigure[$k=4$]{
		\includegraphics[width=1.6in]{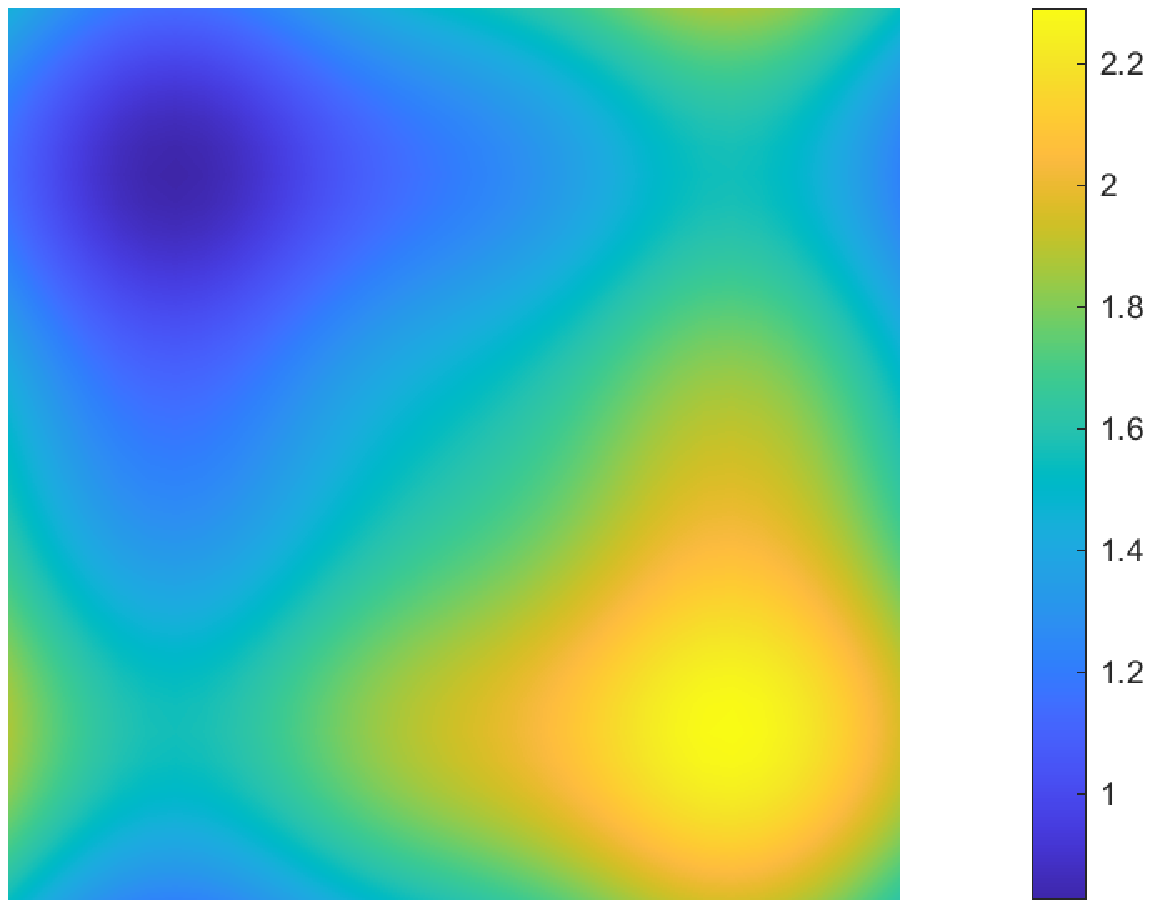}
	}
	\vspace{0em}
	\caption{
		The phase of the x-component of the optimized pattern ${\bm \theta}_k({\bf s})$.
	}
	\label{img:simulation_phase}
	\vspace{-1em}
\end{figure}

	To show the pattern functions optimized by the proposed pattern-division multiplexing, we present the normalized amplitude of the x-component of the optimized pattern functions ${\bm \theta}_k({\bf s})$ for the former four receivers 1-4 in Fig. \ref{img:simulation_amplitude}, and their phase in Fig. \ref{img:simulation_phase}, respectively. From these two figures, it is interesting to observe that, after designing the patterns (i.e., the current distributions) via the proposed pattern-division multiplexing scheme, the patterns ${\bm \theta}_k({\bf s})$ for different receivers are nearly orthogonal. In particular, as shown in Fig. \ref{img:simulation_amplitude}, the power of the patterns that carry different symbols are distributed in non-overlapping regions, in this way, the inter-receiver interference at each electromagnetic-wave receiver can be well eliminated.
	Besides, from Fig. \ref{img:simulation_phase} one can notice that, the phase of the patterns carrying different symbols are symmetrically distributed, which means that the electromagnetic-waves for the four receivers are radiated towards four different spatial directions and focused on the four receivers respectively. This interesting phenomenon is similar to the design result of \ac{mimo} beamforming, which aims to generate orthogonal beams towards multiple receivers. It also intuitively shows the reason why the proposed pattern design scheme can improve the sum-rate. We can conclude that, both of these two figures have provided intuitive explanations for the sum-rate improvement, which have demonstrated the effectiveness of the proposed pattern-division multiplexing technique\footnote{More simulation results and insights can be found in the journal version of this paper \cite{Zijian'22}.}.

\section{Conclusions}\label{sec:con}
In this paper, we proposed the pattern-division multiplexing to fill in the gap of a flexible pattern design scheme for \ac{cap-mimo}. Specifically, we first derived the system model of \ac{cap-mimo}, which allowed us to formulate the sum-rate maximization problem. Then, we proposed a general pattern-division multiplexing technique to flexibly design the patterns of \ac{cap-mimo}. By applying series expansion to project the continuous functions of the patterns onto an orthogonal basis space, the design of continuous pattern functions were transformed to the design of their projection lengths on finite orthogonal bases. Based on this technique, we proposed a pattern design scheme to solve the formulated sum-rate maximization problem. Simulation results showed that, the patterns designed by the proposed scheme for multiple receivers were almost mutually orthogonal. Particularly, the sum-rate achieved by the proposed scheme was about 260\% higher than that achieved by the existing wavenumber-division multiplexing scheme, which demonstrated the effectiveness of the proposed scheme.

\footnotesize
\bibliographystyle{IEEEtran}
\bibliography{IEEEabrv,reference}
\end{document}